\documentclass[preprint]{elsarticle}
\usepackage{feynmf,amsmath,amssymb}
\usepackage{graphicx,graphics}
\usepackage{slashed}
\usepackage{subfig}
\begin{document}

\begin{fmffile}{upgm2fmf}

\def\U{\mathcal{U}}
\def\sU{\slashed{\mathcal{U}}}
\def\O{\mathcal{O}}
\def\d{d_{\mathcal{U}}}
\def\Re{{\rm Re}}
\def\Im{{\rm Im}}
\def\wp{\tilde{\Delta}_\sU}
\def\disp{{\rm disp}}
\def\tev{\,{\ifmmode\mathrm {TeV}\else TeV\fi}}
\def\gev{\,{\ifmmode\mathrm {GeV}\else GeV\fi}}

\def\Acknowledgements{\bigskip  \bigskip \begin{center} \begin{large}
             \bf ACKNOWLEDGEMENTS \end{large}\end{center}}

\title{
\begin{flushright}
\mbox{\normalsize SLAC-PUB-13468}
\end{flushright}
\vskip 15 pt
Vector Unparticle Contributions to Lepton g-2}
\author{J. A. Conley and J. S. Gainer}
\address{SLAC National Accelerator Laboratory, 2575 Sand Hill Road, Menlo Park, CA, 94025}
\date{November 25, 2008}

\begin{abstract}
The generic unparticle propagator may be modified in two ways.
Breaking the conformal symmetry effectively adds a mass term to the
propagator, while considering vacuum polarization corrections adds a
width-like term.  Both of these modifications result naturally from
the coupling of the unparticle to standard model (SM) fields.  We
explore how these modifications to the propagator affect the
calculation of the lepton anomalous magnetic moment using an integral
approximation of the propagator that is accurate for $d\lesssim1.5$,
where $d$ is the unparticle dimension.  We find that for this range of
$d$ and various values of the conformal breaking scale $\mu$, the
value of $g-2$ calculated when allowing various SM fermions to run in
the unparticle self-energy loops does not significantly deviate from
the value of $g-2$ when the width term is ignored.  We also
investigate the limits on a characteristic mass scale for the
unparticle sector as a function of $\mu$ and $d$.
\end{abstract}

\maketitle

\section{Introduction}

Unparticle physics \cite{Georgi:2007ek,Georgi:2007si}
represents an intriguing possibility for physics beyond the standard
model.  
In this scenario, there is a new sector which is conformal below a
mass scale $\Lambda_{\U}$.  It can be described by an effective field theory
of so-called ``unparticle'' operators, which in some respects imitate
a non-integer number of massless particles.
The phenomenology of unparticles has been studied in great detail.
In particular, the anomalous magnetic moments of the electron and the
muon have been calculated in various unparticle
scenarios~\cite{Cheung:2007zza,Hektor:2008xu,Hur:2007cr,Liao:2007bx,Luo:2007bq}.
Here we explore the effect of adding a width-like term to the denominator
of the propagator on the calculation of electron and muon $g-2$.  We also
study in detail the relationship between the conformal breaking scale $\mu$
and these quantities.

The propagator for unparticles can be obtained from the spectral 
representation
\begin{equation}\label{dispersion}
 P_{\U}=
\int d^4x~ e^{ipx}\langle \O_{\U}(x)
\O^\dagger_{\U}(0)
\rangle=\int_0^\infty\frac{dM^2}{2\pi}\rho_\U(M^2)
\frac{i}{p^2-M^2+i\epsilon},
\end{equation}
where $\O_\U$ is an unparticle operator and $\rho_\U(M^2)$ is the
unparticle spectral function.
If $O_\U$ has scaling dimension $d$, conformal invariance
determines 
\begin{equation}\label{spec}
  \rho_\U(M^2)=A_{d}(M^2)^{d-2}\;,
\end{equation}
where
\begin{equation}
A_{d}=\frac{16\pi^{5/2}}{(2\pi)^{2d}}\frac{\Gamma(d+1/2)}
{\Gamma(d-1)\Gamma(2d)}
\end{equation}
is chosen so that the $d\to 1$ limit gives the propagator for a
normal particle.  Following~\cite{Georgi:2007si} we will take
$d\in(1,2)$.
Inserting this spectral function into Eq.~\ref{dispersion}~and
evaluating the integral yields the propagator
\begin{equation}\label{prop}
   P_\U(p^2)=
    \frac{iA_{d}}{2\sin(\pi d)}(-p^2-i\epsilon)^{d-2}\;.
\end{equation}
The above is valid for a scalar unparticle.  For vector unparticles,
which we consider in this paper, this propagator must be
multiplied by the projector $-(g_{\mu\nu}+bp^\mu p^\nu/p^2)$.  The
constant $b$ is in principle undetermined because the unparticle
theory need not have a gauge symmetry, but it will not affect the
results of any of our calculations so we set it to zero from now on.
The propagator is then $P_\U^{\mu\nu}(p^2)=-g^{\mu\nu}P_\U(p^2)$.

\begin{figure}%
  \centering%
  \subfloat[][]{%
    \includegraphics[width=5.5cm]{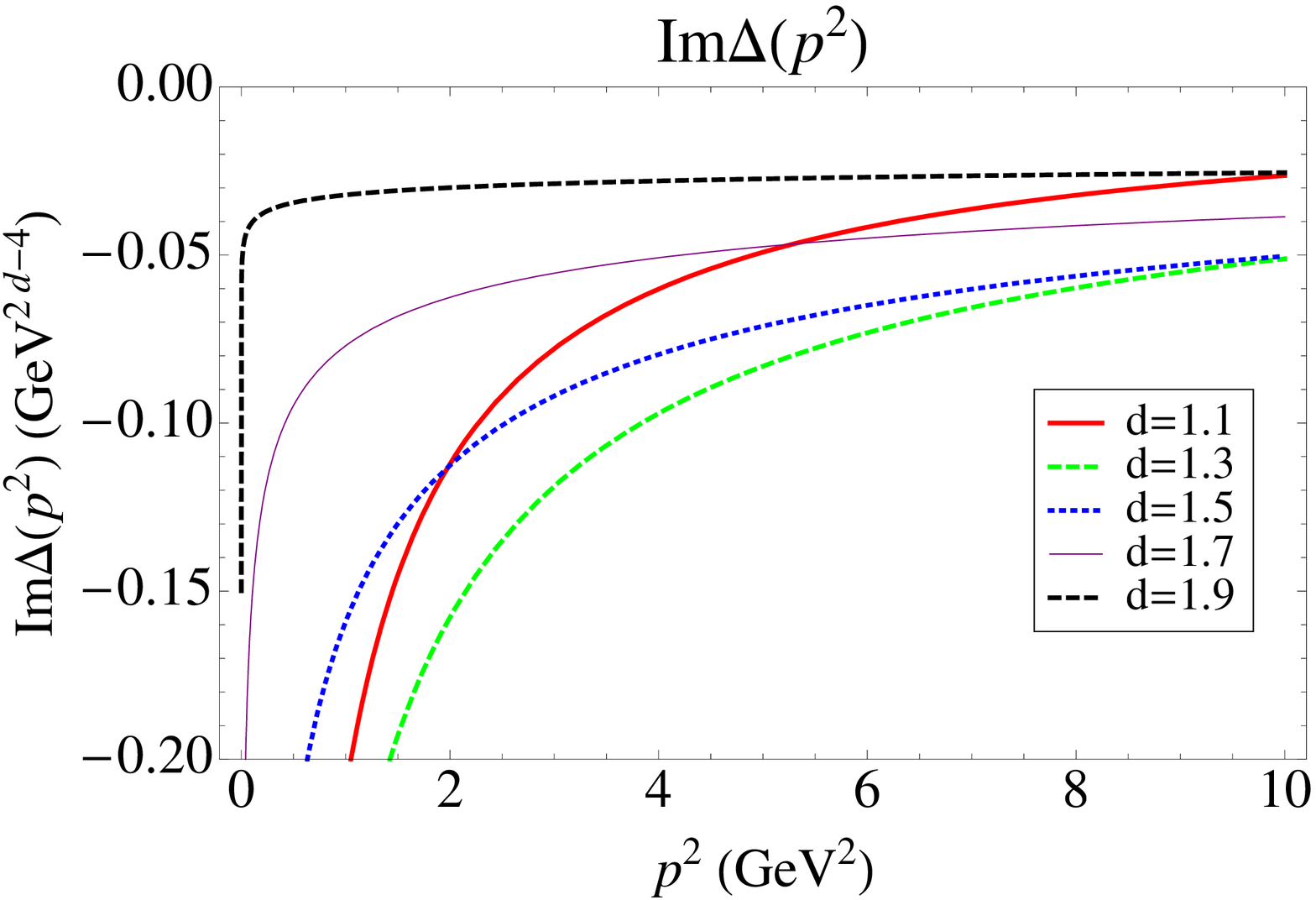}%
    \label{fig:ImP}%
  }%
  \qquad%
  \subfloat[][]{%
    \includegraphics[width=5.5cm]{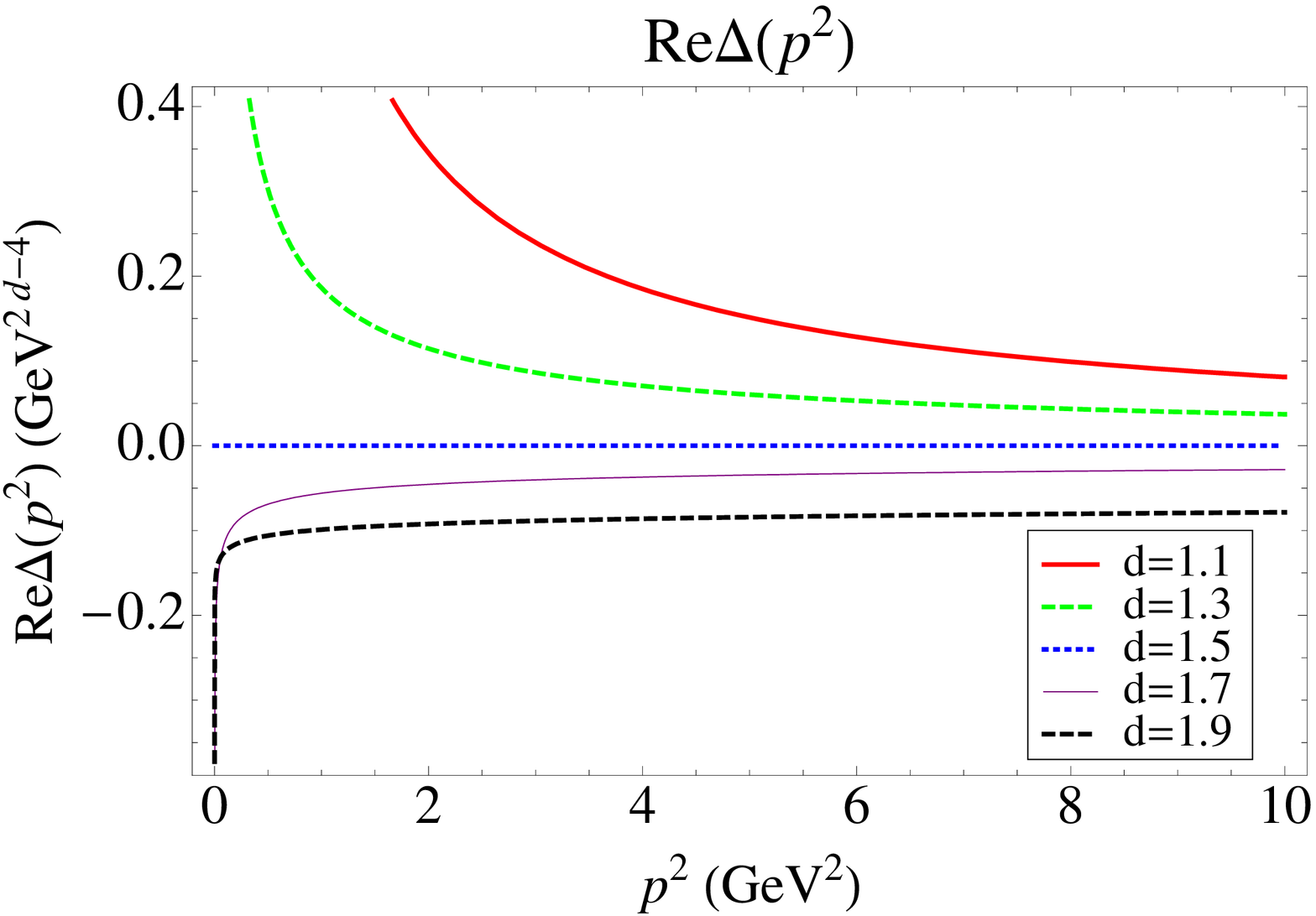}%
    \label{fig:ReP}%
  }%
  \caption{The imaginary part (left panel) and real part (right panel) of the unparticle propagator
     $\Delta_\U(p^2)$ for various values of unparticle dimension $d$,
     in the absence of unparticle sector conformal symmetry breaking.}%
\end{figure}

To get a feel for the behavior of this unusual propagator, we show the
imaginary and real parts of $\Delta_\U(p^2)=-iP_\U(p^2)$ for positive
$p^2$ in Figs.~\ref{fig:ImP} and \ref{fig:ReP}, respectively.  It will be
convenient for us to use $\Delta_\U$ rather than $P_\U$ in our analysis
and we will often refer to $\Delta_\U$ as ``the propagator.'' It should be
noted that the imaginary part of $\Delta_\U(p^2)$ is zero for negative
$p^2$, while the real part is not.



Following~\cite{Bander:2007nd} we shall write the coupling of vector
unparticles to Standard Model (SM) fermions in the form 
\begin{equation}\label{coupling}
  \Delta{\mathcal{L}}=i\frac{c}{\Lambda_3^{d-1}}\bar{\psi}
\gamma_\mu\psi\O^\mu_\U\;.
\end{equation}
The parameter $\Lambda_3$ is not fundamental; it can be expressed as
the combination
\begin{equation}
  \Lambda_3 = \bigg(\frac{\Lambda_\U}{M}\bigg)^
         {\frac{d_{UV}-1}{d-1}}\Lambda_\U\;,
\end{equation} 
where $M$ and $d_{UV}$ are parameters of the underlying UV theory
as described in~\cite{Bander:2007nd}.
Since the coupling $c$ and the scale $\Lambda_3$ only appear in the
combination given in Eq.~\ref{coupling}, and we have no theoretical
guidance as to what $c$ may be (other than naturalness, which might
suggest that $c$ is $\O(1)$), in our analysis we will take
$c=1$ and let the coupling strength vary by varying $\Lambda_3$.

\begin{figure}%
  \centering%
  \subfloat[][]{%
    \includegraphics[width=5.5cm]{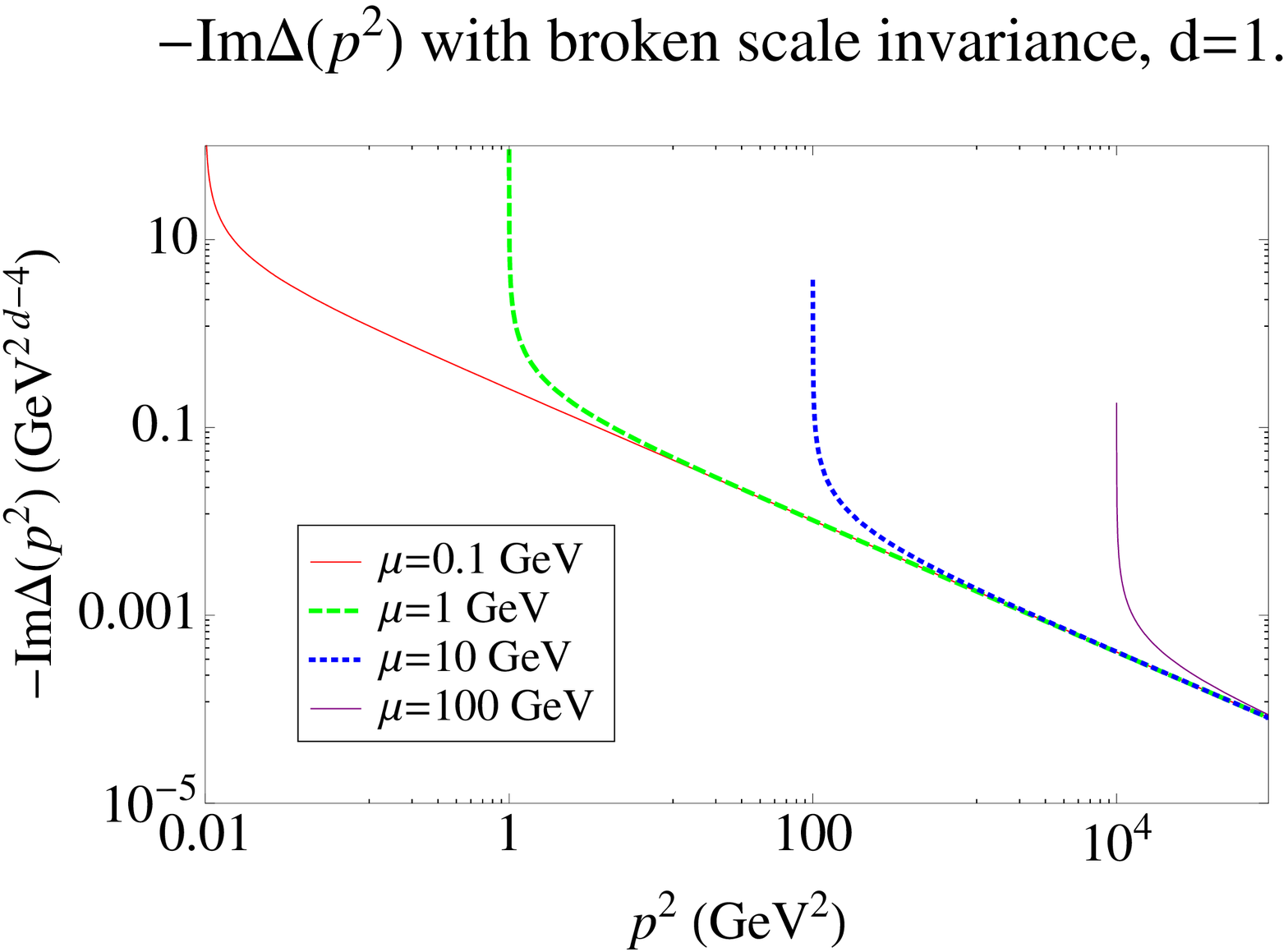}%
    \label{fig:ImPmu}%
  }%
  \qquad%
  \subfloat[][]{%
    \includegraphics[width=5.5cm]{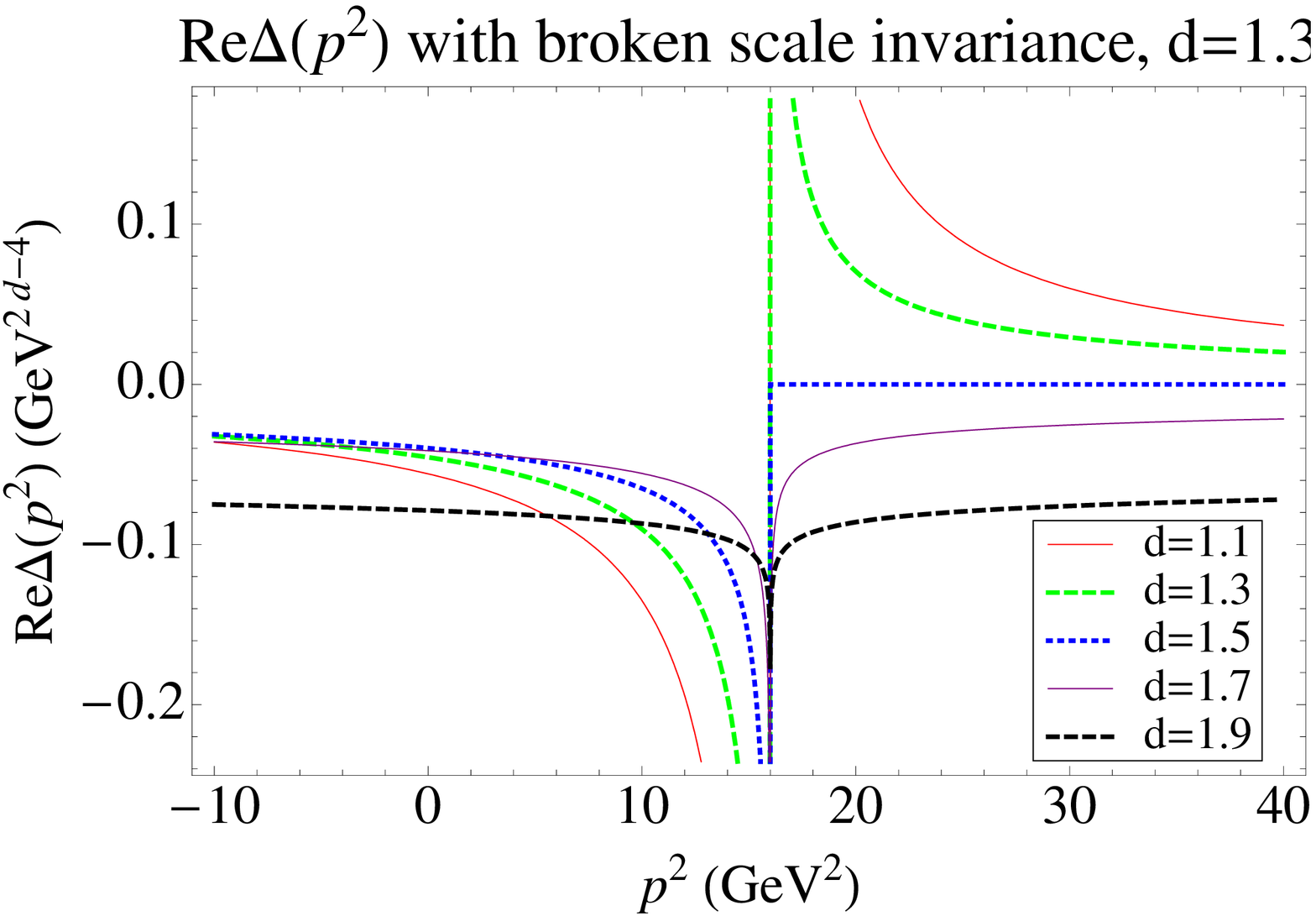}%
    \label{fig:RePmu}%
  }%
  \caption{The negative imaginary part (left panel) and real part (right panel) of the unparticle propagator with broken
     conformal invariance $\Delta_\sU(p^2)$ for $d=1.3$.  The negative imaginary part is 
     shown for four different values of the conformal breaking scale $\mu$, while the real part is shown for five different 
     values of the unparticle dimension $d$.}%
\end{figure}

An interesting modification of the unparticle scenario involves breaking the
conformal symmetry of the unparticle sector at a scale $\mu$.  This can
result from interactions between the unparticle and the Higgs, for example.
An ansatz for the spectral function in this case essentially consists in
removing modes below $\mu^2$ from the spectrum, yielding~\cite{Fox:2007sy} 
\begin{equation}\label{broken spec}
\rho_\sU(M^2)=A_{d}\theta(M^2-\mu^2)(M^2-\mu^2)^{d-2}\;.
\end{equation}
With this spectral function, the propagator becomes
\begin{equation}\label{brokenprop}
   P_\sU(p^2)=
    \frac{iA_{d}}{2\sin(\pi d)}(-(p^2-\mu^2)-i\epsilon)^{d-2}\;.
\end{equation}
For the fractional power, we take the standard branch cut such that
\begin{equation}
  (\mu^2-p^2-i\epsilon)^{2-d}=\begin{cases}
    (\mu^2-p^2)^{2-d} & \text{if}~\mu^2\geq p^2,\\
    (p^2-\mu^2)^{2-d}e^{i\pi d} & \text{if}~p^2>\mu^2\;.
  \end{cases}
\end{equation}
In Fig.~\ref{fig:ImPmu}, we show the imaginary part of
$\Delta_\sU(p^2)=-iP_\sU(p^2)$ for $d=1.3$ and different
choices of $\mu$.  It rises steeply near $p^2=\mu^2$ and vanishes for
$p^2<\mu^2$.  
In Fig.~\ref{fig:RePmu} we display $\Re\Delta_\sU(p^2)$ for
$\mu=4~\gev$ for different values of $d$.  Unlike the 
imaginary part, the real part is nonzero for $p^2<\mu^2$, though for
$d=1.5$ it is zero for $p^2>\mu^2$.

Some of the strongest experimental
constraints on unparticle physics can be evaded with a sufficiently
large value of $\mu$~\cite{Fox:2007sy,Barger:2008jt}.  To understand this, note that when there is 
a nonzero $\mu$, the form of the spectral function (see Eq.~\ref{broken spec})
is such that low $M^2$ modes no longer contribute to the propagator
(cf. Eq.~\ref{dispersion}).  It is well-known that the bulk of the
contribution to $g-2$ in QED comes from low $M^2$ modes in the
propagator~\cite{Berestetsky,McClure:1964ne,Drell:1965hg,Brodsky:1980zm}; hence, for large enough values of $\mu$,
the contribution to $g-2$ from unparticles is negligible.
It has been suggested~\cite{Barger:2008jt}
that the propagator given in Eq.~\ref{brokenprop} should be
further modified by summing the contributions to the propagator from
diagrams containing an arbitrary number of vacuum polarization
bubbles, as is done when finding the renormalized mass and width of a
``normal'' particle.  It will be interesting to see how this modification
to the propagator affects the above intuition regarding the size of $g-2$.
In particular, we wish to know whether such a modification, which should lead to a nonzero
spectral function for values of $M^2$ less than $\mu^2$, will result in 
a significantly larger value of $g-2$.

In this paper we calculate the contribution to muon $g-2$ from
a vector unparticle both with and without vacuum polarization
corrections to the propagator.  We obtain bounds on $d$, $\mu$,
and $\Lambda_3$ from the experimental limits on muon and 
electron $g-2$, and investigate how these limits are affected 
by these corrections.
Specifically we will consider the modification to the unparticle 
propagator from vacuum polarization graphs with fermions running in 
the loop.  In particular, we will consider the following 3 cases:
\begin{enumerate}
\item The unparticle couples only to the lepton for which we are
  calculating $g-2$.
\item The unparticle couples to all leptons with equal
  strength, but to no other particles.
\item The unparticle couples to all standard model fermions
 with equal strength, but to no other particles.
\end{enumerate}
The calculation of the unparticle contribution to $g-2$ in these cases
will require obtaining an approximate dispersion
integral representation of the modified propagator, whose validity and
accuracy we will investigate.

\section{Vacuum polarization corrections to the propagator}

We will first consider the modification of the propagator
from vacuum polarization diagrams.  While in principle we could
consider decays to scalar or vector particles, and possibly even other
species of unparticle, for simplicity we confine our attention to loops involving
standard model fermions.  For a given fermion in the loop, the amplitude 
for the vacuum polarization graph is 
\begin{equation}\label{vacuum polarization}
   \parbox{80pt}{
      \begin{fmfgraph*}(80,50)
        \fmfleft{i}
        \fmfright{o}
        \fmflabel{$q$}{i}
        \fmf{dbl_wiggly}{i,v1}
        \fmf{dbl_wiggly}{v2,o}
        \fmf{fermion,left}{v1,v2,v1}
        \fmfforce{(.3w,.5h)}{v1}
        \fmfforce{(.7w,.5h)}{v2}
      \end{fmfgraph*}
    } 
  \quad =\; i(g^{\mu\nu}-q^\mu q^\nu/q^2)\Pi(q^2)\;,
\end{equation}
where
\begin{equation}
  \Pi(q^2)=-\frac{1}{2\pi^2}\left(\frac{c}{\Lambda_3^{d-1}}\right)^2q^2B
  \left(\frac{q^2}{4m_f^2}\right)\;,
\end{equation}
and $B(\beta)$ is the Feynman parameter integral
\begin{equation}
  B(\beta)=\int_0^1dx\,x(1-x)\log(1/(1-4\beta x(1-x)))\;.
\end{equation}
Our unparticle theory is not a gauge theory. Since the rest of
the g-2 calculation involves only QED, however, the
contribution proportional to $q^\mu q^\nu$ will vanish by the Ward
identity.

\begin{figure}%
  \centering%
  \subfloat[][]{%
    \includegraphics[width=5.5cm]{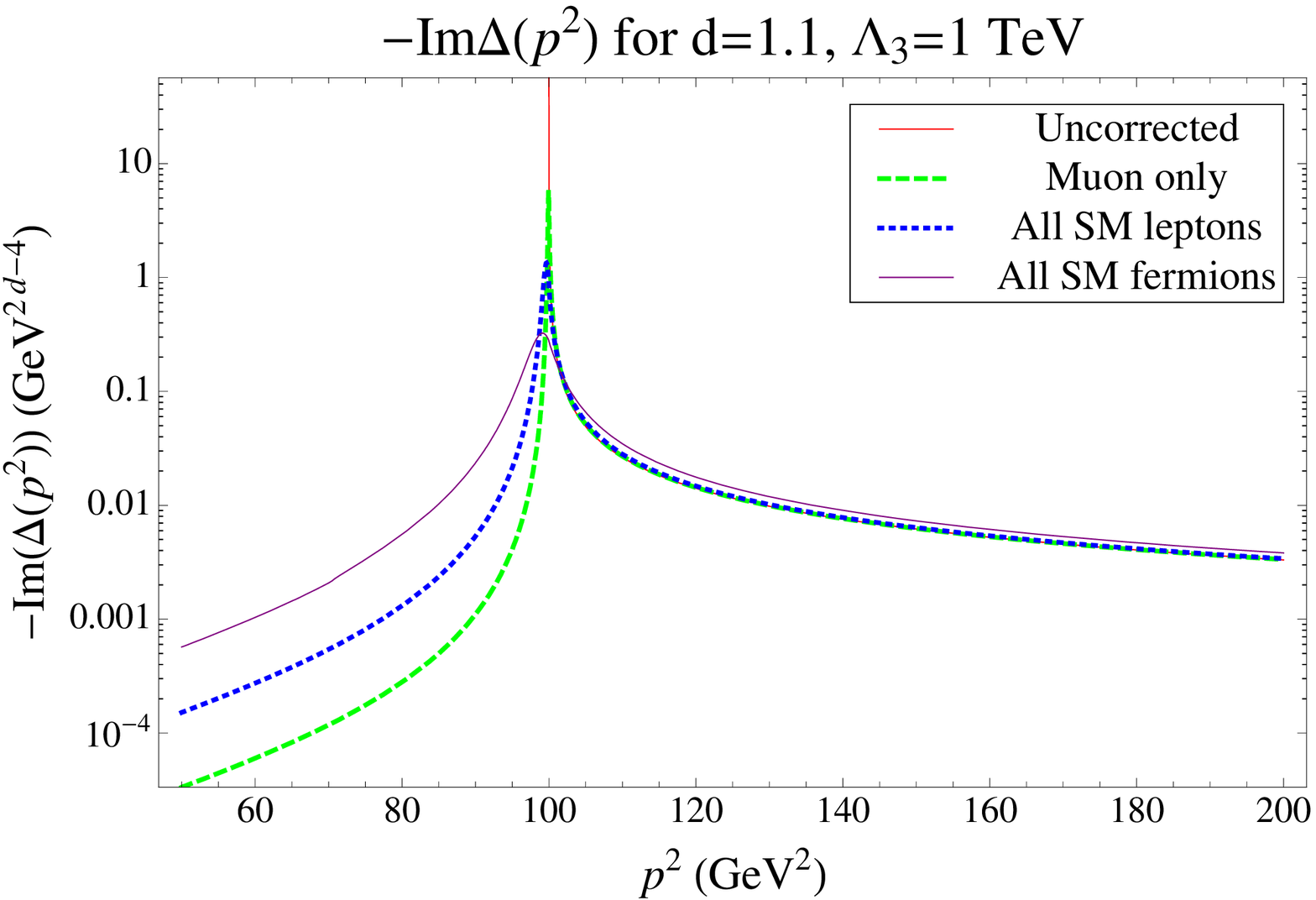}%
    \label{fig:ImPw}%
  }%
  \qquad%
  \subfloat[][]{%
    \includegraphics[width=5.5cm]{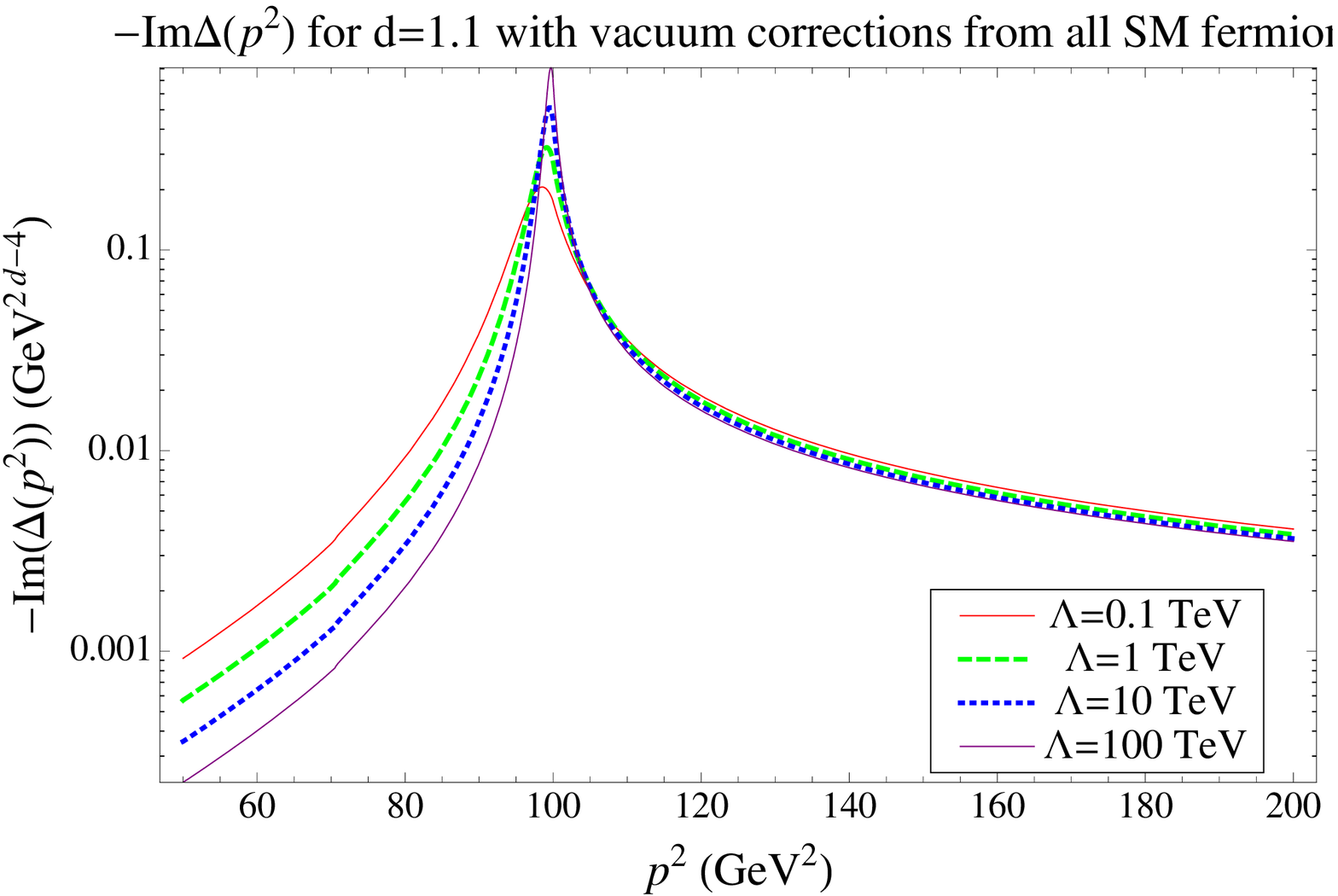}%
    \label{fig:ImPwl}%
  }%
  \caption{The negative imaginary part of the
    propagator, for $d=1.1$ and $\mu=10~\gev$, for  $\Lambda_3 = 1$ TeV and different
    choices for loop corrections (left panel) and for various values of $\Lambda_3$ (right panel).
    In the left panel, the ``uncorrected'' curve shows the propagator in
    the absence of vacuum polarization corrections, while the other three curves show
    the value of the propagator if only muons, all leptons, or
    all SM fermions are allowed to run in the loops of the
    self-energy terms which lead to a width-like term in the unparticle
    propagator.  In the right panel, different curves
    represent different values of $\Lambda_3$; since this parametrizes
    the fermion-unparticle coupling, it affects the shape as well as the
    normalization of the propagator if the propagator includes the
    width-like factor from vacuum polarization corrections.}%
\end{figure}

Summing over diagrams with an arbitrary number of vacuum polarization
loop insertions yields the propagator 
$\tilde{P}^{\mu\nu}_{\sU}(p^2)=-g^{\mu\nu}\tilde{P}_{\sU}(p^2)$, with
\begin{equation}\label{wprop}
  \tilde{P}_{\sU}(p^2)=\frac{iA_d}{2\sin(\pi d)
(\mu^2-p^2-i\epsilon)^{2-d}-A_d\Pi(p^2)}\;.
\end{equation}
To see the effect of this modification to the propagator, in Fig.~\ref{fig:ImPw} we show 
$-\Im\wp(p^2)=-\Im(-i\tilde{P}_\sU(p^2))$ for $d=1.1$, $\mu=10~\gev$, and $\Lambda_3=1~\tev$ 
with four different choices of vacuum polarization corrections.  The peak near $p^2=\mu^2$
is broadened considerably by the corrections, showing that the $A_d\Pi(p^2)$ term in the
denominator of the propagator has an effect similar to that of a decay width for a normal particle.
In Fig.~\ref{fig:ImPwl}, on the other hand, we show 
$-\Im\wp(p^2)$ again for $d=1.1$ and $\mu=10~\gev$, but now we only consider vacuum polarization corrections
from all SM fermions and let $\Lambda_3$ take on four different values.  Because decreasing $\Lambda_3$
effectively increases the unparticle-SM coupling, the propagator peak is broadened more for lower $\Lambda_3$.

\section{Calculating $g-2$ using a dispersion integral}

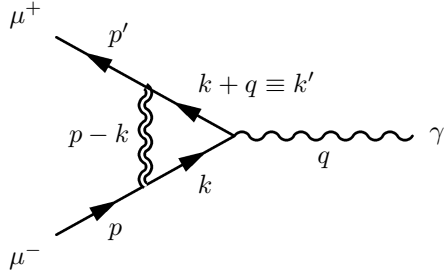
\begin{figure}
  \begin{center}
    \begin{fmfgraph*}(150,75)
      \fmflabel{$\mu^+$}{i1}
      \fmflabel{$\mu^-$}{i2}
      \fmflabel{$\gamma$}{o1} 
      \fmfleft{i2,i1}
      \fmfright{o1}
      \fmf{fermion,label=$p$,l.side=right}{i2,v3}
      \fmf{fermion,label=$k$,l.side=right}{v3,v1}
      \fmf{fermion,label=$k+q\equiv k'$}{v1,v2}
      \fmf{fermion,label=$p'$}{v2,i1}
      \fmf{dbl_wiggly,tension=0,label=$p-k$,l.side=left}{v3,v2}
      \fmf{boson,label=$q$}{v1,o1}
    \end{fmfgraph*}
  \end{center}
  \caption{The unparticle correction to the muon $g-2$.  The photon
    momentum $q$ is incoming and the muon momenta follow the fermion
    lines.}
  \label{fig:loop}
\end{figure}

Now that we have the form for the unparticle propagator with vacuum polarization
corrections, we turn our attention to the question of how to calculate 
the one-loop contribution to $g-2$ from unparticles both with and without these corrections to 
the propagator.
This contribution comes from
the diagram in Fig.~\ref{fig:loop}.
The amplitude $i\mathcal{M}^\mu$ corresponding to this diagram satisfies
\begin{equation}\label{g2int}
  i\mathcal{M}^\mu\propto\int\frac{d^4k}{(2\pi)^4}
  \frac{\bar{u}(p')(\slashed{k}\gamma^\mu\slashed{k}'+m^2\gamma^\mu-2m(k+k')^\mu)u(p)}
  {(k^2-m^2+i\epsilon)(k'^2-m^2+i\epsilon)D_\U}\;,
\end{equation}
where $D_\U$ is given by
\begin{equation}\label{d1}
  D_\U= (\mu^2-(p-k)^2-i\epsilon)^{2-d}
\end{equation}
for an unparticle propagator without vacuum polarization corrections, and 
\begin{equation}\label{d2}
 D_\U=(\mu^2-(p-k)^2-i\epsilon)^{2-d}-\frac{A_d}{2\sin(\pi d)}\Pi((p-k)^2)
\end{equation}
for an unparticle propagator with vacuum polarization corrections.

Note that the numerator is identical to that which one obtains in the
calculation of the QED one-loop contribution to $g-2$.
Since we are only considering vector unparticles, which conserve parity, we can write the amplitude as
\begin{equation}
  i\mathcal{M}^\mu=ie\bar{u}(p')\left(\gamma^\mu F_1(q^2)+i\frac{\sigma^{\mu\nu}}{2m}F_2(q^2)\right)u(p)\;.
\end{equation}
The unparticle contribution to the quantity $a=g/2-1$ is then given by $a^\U=F_2(0)$.

When calculating the QED one-loop contribution to $g-2$ (where the particle in the loop is a photon rather
than an unparticle, and hence $D_\U\to (p-k)^2+i\epsilon$), we can use the method of Feynman parameters to combine the three factors
in $D$ and shift the integration variable to $l=k+yq-zp$.  The result is 
\begin{equation}\label{fpsm}
  F_2(0)=8ie^2m^2\int\frac{d^4l}{(2\pi)^4}\int_0^1dx\,dy\,dz\frac{z(1-z)}{(l^2-\Delta+i\epsilon)^3}\;.
\end{equation}
Note that there are no powers of $l$ in the numerator of the integrand.  As a result, the integral
is convergent.  The simplifications (such as determining that terms proportional to $l^2$
contribute to $F_1$ rather than $F_2$) that make this explicit require the use of Feynman parameters.

In the case of unparticles without vacuum polarization corrections to the propagator, the Feynman parameter method
can still be used, though in this case the denominator is raised to the power $4-d$.  With the corrected propagator, 
however, it is impossible to combine the terms in the denominator
into an expression quadratic in momentum variables.  Hence, we cannot use Feynman parameters in this case 
to obtain a relatively simple, manifestly convergent
integral expression for $F_2(0)$ analogous to Eq.~\ref{fpsm}.  It is difficult to imagine that one can calculate the
integral in Eq.~\ref{g2int} analytically in this case.  One could hope to perform the integral numerically, but without 
the simplifications obtained using Feynman parameters, the integral has divergent contributions whose cancellation
would be difficult to guarantee in a numerical calculation. We therefore seek to recast the integral in a form amenable
to the use of Feynman parameters. 

To do this, we try writing the unparticle propagator as a dispersion integral 
(cf. Eq.~\ref{dispersion}) which contains the standard vector boson propagator. 
Because $\rho(M^2)$ is independent of the loop momentum, it can be pulled
out of the loop momentum integral.  The vector boson propagator, however, remains 
under the loop integral and we end up with
\begin{equation}
  \label{dispa}
  a^\U= \int_0^\infty \frac{dM^2}{2\pi}\rho(M^2) I(M^2),
\end{equation}
where $\rho(M^2)$ is the spectral function and $I(M^2)$ is the
contribution to $g-2$ at one loop from a vector boson with mass $M$
and coupling to fermions given by Eq.~\ref{coupling}.  
Specifically we have that $I(M^2)$ is given by the Feynman parameter
integral
\begin{equation}\label{first IM2}
  I(M^2) = \frac{c^2}{4\pi^2\Lambda_3^{2-2d}}\int_0^1 dx
  \frac{x^2(1-x)^2 m^2}{x^2m^2 + (1-x) M^2}\;,
\end{equation}
where $m$ is the mass of the fermion for which we are calculating
$g-2$ (generally the muon), and as noted above we will set c=1 in what follows.
The Feynman parameter integral in Eq.~\ref{first IM2}~can be performed
to obtain an
analytical expression for $I(M^2)$, leaving the $M^2$ integral in
Eq.~\ref{dispa}~to be done numerically.

This method of expressing the propagator as a dispersion integral 
clearly works in the case without vacuum polarization corrections, since we already have
expressions given above for $\rho(M^2)$ (Eqs.~\ref{spec} and \ref{broken spec}).
Of course in this case, as mentioned above, one could do the calculation 
using Feynman parameters \cite{Cheung:2007zza,Liao:2007bx} without recourse to the dispersion integral.  
This provides a useful check on our results.  

To use this method when we include vacuum polarization corrections to the
unparticle propagator, however, we still need to express that propagator
as a dispersion integral.  In the next section, we will study whether this can be done.

\section{Expressing the propagator as a dispersion integral}

We need to find an expression for 
the spectral function of the vacuum polarization-corrected unparticle propagator.
To this end, we first note that the 
spectral function for the ``tree-level'' propagator in both the
unbroken (Eq.~\ref{spec}) and broken (Eq.~\ref{broken spec})  conformal
symmetry cases satisfies the relation $\rho(M^2)=2\Im \Delta(M^2)$.
In other words, the spectral function is (twice) the imaginary part of the
propagator.  This is actually due to a general result in complex analysis,
which states that
for any function of a complex variable $f(z)$,
\begin{equation}\label{kk}
  f(z)=-\frac{1}{\pi}\int_{-\infty}^{\infty}\frac{\Im\,f(z)}{z-z'+i\epsilon}dz'\;,
\end{equation}
if $f$ is analytic in the upper half-plane.  (The difference between the $\frac{1}{\pi}$
factor in this equation and the $\frac{1}{2\pi}$ in Eq.~\ref{dispersion} accounts for the factor of $2$ in the 
relation $\rho(M^2)=2\Im \Delta(M^2)$, while the $i$ in the numerator of Eq.~\ref{dispersion} combined with 
the $i$ in $\Delta=\Im\,P$ lead to the lack of a relative sign.)
The uncorrected unparticle propagator is indeed analytic in the upper half-plane in both the broken and 
unbroken conformal invariance cases.

It therefore seems like a simple matter to obtain the spectral function for our
vacuum polarization corrected unparticle propagator, Eq.~\ref{wprop}.  It should
be $\tilde{\rho}_\sU(M^2)=2\Im \wp(M^2)$.  The problem is that it is not clear if
the corrected propagator $\wp(M^2)$ is analytic in the upper half-plane.  If it is not,
this relation may not hold.

We would like to test if this condition holds exactly, indicating that the propagator is
analytic in the upper half-plane.
It may also be possible that the propagator is not analytic in the upper half-plane, but is
close enough to an analytic function that the dispersion integral with $\tilde{\rho}_\sU(M^2)=2\Im \wp(M^2)$
serves as a good approximation to the propagator.  To investigate these possibilities, 
we adopt a numerical strategy.  We will {\it assume} 
$\tilde{\rho}_\U(M^2)=2\Im \wp(M^2)$
as a putative definition for the spectral function associated with the vacuum polarization-corrected unparticle
propagator.  We then want to know if the dispersion relation
\begin{equation}\label{wdisp}
  \wp(p^2)\stackrel{?}{=}-\int_0^\infty\frac{dM^2}{\pi}\Im\wp(M^2)
  \frac{1}{p^2-M^2+i\epsilon}
\end{equation}
holds.  Note that the lower range of integration may be taken to be $0$ because $\Im\wp(p^2)$ is zero for $p^2<0$.
The left-hand side of this equation is the true value of the propagator, which we can simply evaluate using Eq.~\ref{wprop}.
The right hand side is our putative spectral representation for the propagator, and we can compute its value by carrying
out the integral numerically.  We can then compare the two sides.  If they do not agree, we learn that,
for the loop-corrected propagator, the relation
$\tilde{\rho}_\sU(p^2)=2\Im \wp(p^2)$ does not in fact hold (or, equivalently, that Eq.~\ref{kk} does not hold),
because the propagator is not analytic in the upper-half plane.  If the two sides of Eq.~\ref{wdisp} do agree to a high
precision, we can conclude that the propagator is analytic (or at least well-enough approximated by an analytic function) in the upper half-plane, so that
the RHS provides the correct spectral representation (or at least good approximation)
of the true propagator.  If this is the case, we can use $\tilde{\rho}_\sU(p^2)$ as a good approximation to the spectral
function in the $g-2$ calculation.

It should be noted that, following \cite{Delgado:2008gj}, when the mass $m$ of the fermion in the loop satisfies $2m>\mu$, 
there will be a value
of $M^2$ for which the denominator of $\wp(M^2)$ is zero.  This leads to an additional delta function
term in $\tilde{\rho}_\sU(M^2)$. 
Such a value can only occur when the loop contribution $\Pi(M^2)$ and the first term in the denominator of the propagator,
$(2\sin(\pi d)/A_d)(\mu^2-p^2-i\epsilon)^{2-d},$ are both real.  The loop contribution from a given fermion with mass $m$ is real for 
$M^2<4m^2$ while the first term in the denominator of the propagator is real for $M^2<\mu^2$.  Hence, the denominator cannot
vanish for $\mu^2>M^2>4m^2$, where $m$ is the mass of the lightest fermion running in the loop, unless there were some fortuitous
cancellation of the imaginary parts of the loop contributions from the various fermion species.  It can be easily verified
that there is no such cancellation.  It may be further shown that, given the signs of the imaginary parts of the various
terms in the denominator of the propagator, it also cannot have a zero for $M^2>\mu^2$.  For $M^2<4m^2$, where $m$ is the mass
of the lightest fermion running in the loop, both terms in the denominator are strictly real.  In the situations we consider, 
however, where the lightest fermion in the loop is either an electron or a muon, and $\mu$ takes values of 10 or 100~\gev, the
absolute value of the first term in the propagator, in this region, is always much larger than the absolute value of the
loop correction term, so no cancellation can occur.  Thus, in the situations we consider, there are no delta function
contributions to the unparticle spectral function.


In Table~\ref{tbl}, 
\begin{table}[htbp]
  \small
  \begin{center}
    \begin{tabular}{|c|c|c|c|c|c|}
      \hline
      $d,~\mu,~p^2$ & Value & None & Muon & Leptons & All \\ \hline
      $ 1.2, 10, 5 $ & $-1.2\times 10^{-2} $ & $ -8.\times 10^{-14} $ & $ -1.3\times 10^{-8} \
      $ & $ -1.2\times 10^{-7} $ & $ -1.7\times 10^{-5} $ \\ \hline
      
      $ 1.2, 10, 500 $ & $3.1\times 10^{-3} $ & $ 6.6\times 10^{-15} $ & $ 4.\times 10^{-9} $ \
      & $ 5.4\times 10^{-7} $ & $ 6.7\times 10^{-5} $ \\ \hline

      $ 1.2, 10, 50000 $ & $6.5\times 10^{-5} $ & $ -1.1\times 10^{-13} $ & $ \
      1.9\times 10^{-7} $ & $ 2.2\times 10^{-5} $ & $ 3.2\times 10^{-3} $ \\ \hline

      $ 1.2, 100, 5 $ & $-2.9\times 10^{-4} $ & $ 6.5\times 10^{-12} $ & $ -3.3\times 10^{-8} \
      $ & $ -2.1\times 10^{-6} $ & $ -2.1\times 10^{-4} $ \\ \hline

      $ 1.2, 100, 500 $ & $-3.\times 10^{-4} $ & $ 6.6\times 10^{-12} $ & $ \
      -2.1\times 10^{-8} $ & $ -2.\times 10^{-6} $ & $ -2.1\times 10^{-4} $ \\ \hline

      $ 1.2, 100, 50000 $ & $7.8\times 10^{-5} $ & $ -1.1\times 10^{-13} $ & $ \
      8.3\times 10^{-8} $ & $ 7.9\times 10^{-6} $ & $ 8.4\times 10^{-4} $ \\ \hline

      $ 1.4, 10, 5 $ & $-1.4\times 10^{-2} $ & $ -1.3\times 10^{-15} $ & $ -7.3\times 10^{-6} \
      $ & $ -4.9\times 10^{-5} $ & $ -3.8\times 10^{-4} $ \\ \hline

      $ 1.4, 10, 500 $ & $1.9\times 10^{-3} $ & $ 4.6\times 10^{-13} $ & $ 5.6\times 10^{-5} \
      $ & $ 3.7\times 10^{-4} $ & $ 2.9\times 10^{-3} $ \\ \hline

      $ 1.4, 10, 50000 $ & $1.\times 10^{-4} $ & $ -3.4\times 10^{-13} $ & $ 1.\times 10^{-3} \
      $ & $ 6.8\times 10^{-3} $ & $ 5.6\times 10^{-2} $ \\ \hline

      $ 1.4, 100, 5 $ & $-8.8\times 10^{-4} $ & $ 4.2\times 10^{-13} $ & $ -8.2\times 10^{-5} \
      $ & $ -4.9\times 10^{-4} $ & $ -3.4\times 10^{-3} $ \\ \hline

      $ 1.4, 100, 500 $ & $-9.1\times 10^{-4} $ & $ 4.3\times 10^{-13} $ & $ \
      -8.\times 10^{-5} $ & $ -4.8\times 10^{-4} $ & $ -3.3\times 10^{-3} $ \\ \hline

      $ 1.4, 100, 50000 $ & $1.2\times 10^{-4} $ & $ -2.1\times 10^{-13} $ & $ \
      6.1\times 10^{-4} $ & $ 3.7\times 10^{-3} $ & $ 2.6\times 10^{-2} $ \\ \hline

      $ 1.6, 10, 5 $ & $-1.9\times 10^{-2} $ & $ 9.\times 10^{-16} $ & $ -9.7\times 10^{-4} $ \
      & $ -2.3\times 10^{-3} $ & $ -6.\times 10^{-3} $ \\ \hline

      $ 1.6, 10, 500 $ & $-3.3\times 10^{-3} $ & $ -4.6\times 10^{-13} $ & $ \
      -5.6\times 10^{-3} $ & $ -1.3\times 10^{-2} $ & $ -3.4\times 10^{-2} $ \\ \hline

      $ 1.6, 10, 50000 $ & $-4.9\times 10^{-4} $ & $ 3.4\times 10^{-13} $ & $ \
      -3.8\times 10^{-2} $ & $ -9.\times 10^{-2} $ & $ -2.3\times 10^{-1} $ \\ \hline

      $ 1.6, 100, 5 $ & $-3.\times 10^{-3} $ & $ -4.4\times 10^{-7} $ & $ -5.\times 10^{-3} $ \
      & $ -1.1\times 10^{-2} $ & $ -2.8\times 10^{-2} $ \\ \hline

      $ 1.6, 100, 500 $ & $-3.1\times 10^{-3} $ & $ -4.6\times 10^{-7} $ & $ \
      -4.9\times 10^{-3} $ & $ -1.1\times 10^{-2} $ & $ -2.7\times 10^{-2} $ \\ \hline

      $ 1.6, 100, 50000 $ & $-5.3\times 10^{-4} $ & $ 2.1\times 10^{-13} $ & $ \
      -2.8\times 10^{-2} $ & $ -6.3\times 10^{-2} $ & $ -1.6\times 10^{-1} $ \\ \hline

      $ 1.8, 10, 5 $ & $-3.5\times 10^{-2} $ & $ -2.7\times 10^{-7} $ & $ -4.1\times 10^{-2} \
      $ & $ -5.7\times 10^{-2} $ & $ -8.2\times 10^{-2} $ \\ \hline

      $ 1.8, 10, 500 $ & $-2.1\times 10^{-2} $ & $ -1.1\times 10^{-13} $ & $ \
      -6.8\times 10^{-2} $ & $ -9.4\times 10^{-2} $ & $ -1.3\times 10^{-1} $ \\ \hline

      $ 1.8, 10, 50000 $ & $-8.\times 10^{-3} $ & $ 8.1\times 10^{-14} $ & $ \
      -1.8\times 10^{-1} $ & $ -2.5\times 10^{-1} $ & $ -3.5\times 10^{-1} $ \\ \hline

      $ 1.8, 100, 5 $ & $-1.4\times 10^{-2} $ & $ -1.1\times 10^{-8} $ & $ -9.4\times 10^{-2} \
      $ & $ -1.2\times 10^{-1} $ & $ -1.8\times 10^{-1} $ \\ \hline

      $ 1.8, 100, 500 $ & $-1.4\times 10^{-2} $ & $ -1.2\times 10^{-8} $ & $ \
      -9.3\times 10^{-2} $ & $ -1.2\times 10^{-1} $ & $ -1.7\times 10^{-1} $ \\ \hline

      $ 1.8, 100, 50000 $ & $-8.3\times 10^{-3} $ & $ 4.9\times 10^{-14} $ & $ \
      -1.5\times 10^{-1} $ & $ -2.\times 10^{-1} $ & $ -2.9\times 10^{-1} $ \\ \hline
    \end{tabular}
  \end{center}
  \caption{The fractional error $\delta^{\disp}$ (see text) in the dispersion integral approximation for the unparticle propagator. 
    An entry in the first column gives the choice of parameters used for that row, with $\mu$ given in GeV and $p^2$ in $\gev^2$.
    The second column gives the actual value of the propagator (without vacuum polarization corrections) for that choice
    of parameters. The remaining 
    columns give the fractional error for the dispersion integral approximation for the propagator $\wp^{\disp}(p^2)$  using different
    choices for vacuum polarization corrections.  ``None'' means no vacuum polarization corrections, ``Muon'' means only the muon loop 
    is included, ``Leptons'' means the vacuum polarization corrections from all charged SM leptons are included, and ``All'' means the
    corrections from all charged SM fermions are included.
  }
  \label{tbl}
\end{table}

we compare the actual loop-corrected propagator (Eq.~\ref{wprop}) to its putative spectral representation
(RHS of Eq.~\ref{wdisp}) for a large number of choices for $d$, $\mu$, and $p^2$.  For each choice of these parameters,
we show the fractional error between the spectral integral approximation for the propagator and its true value.  
That error is obtained by first, using {\it Mathematica} \cite{mathematica}, 
performing the numerical integration to evaluate the RHS of Eq.~\ref{wdisp} (which
we'll denote by $\wp^{\rm disp}(p^2)$).  We then compare this to the actual value of the propagator, obtaining the fractional error
\begin{equation}
  \delta^{\rm disp}\equiv\left|\frac{\wp^{\disp}(p^2)-\wp(p^2)}{\wp(p^2)}\right|\;.
\end{equation}
This fractional error is obtained for four cases.  First, for the case of no loop corrections to the propagator.  In this case,
we know the spectral representation with $\tilde{\rho}_\sU(p^2)=2\Im \wp(p^2)$ to be exact, so we expect zero error, and indeed
find errors small enough to be attributed to inaccuracy in the numerical integration (except when the value of the propagator is
exceedingly small, as explained in the caption).  In the last three columns, the fractional
error is shown with three different sets of fermions allowed to run in the loop in the unparticle propagator---muon only, leptons only, and all SM fermions.
We notice that for all these cases and for all values of $d\lesssim1.5$, the fractional 
error is at most a few percent.  Thus for these values of $d$, we can
use  $\tilde{\rho}_\sU(p^2)=2\Im \wp(p^2)$ to
carry out the calculation of the unparticle contribution to $g-2$ with confidence that the result will be accurate.

\section{Results}

We now have a dispersion representation for the unparticle propagator. It is exact in the cases with unbroken and broken conformal
symmetry but without vacuum polarization corrections. In the case with vacuum polarization corrections, it is a good approximation 
for $d\lesssim1.5$.  We can then numerically integrate Eq.~\ref{dispa} to obtain the unparticle contribution to $g-2$.

Doing this, we find, in fact, that the effect of the propagator corrections on the $g-2$ contribution is small.
Table~\ref{tbltwo} shows the relative difference between the unparticle contribution to $g-2$ with and without
the inclusion of vacuum polarization corrections to the unparticle propagator.  The difference is negligible
for all parameter choices for which the dispersion representation of the propagator is accurate.  Therefore, in the
rest of this section, we will discuss results using the uncorrected unparticle propagator, knowing they will still hold
when the vacuum polarization corrections to the propagator are included.\footnote{In fact we are only certain this is
true for $d\lesssim1.5$. Since for $d\gtrsim1.5$, we do not have an accurate
approximation for the spectral function for the unparticle propagator, we technically have not ruled out the possibility that the effects
of loop corrections to the propagator on $g-2$ are large in this case.}
It is possible, of course, that $\Pi(p^2)$ is significantly modified at high $p^2$ by heavy particles whose presence
we have not taken into account, for example from other hidden sectors to which the unparticle is coupled.  We note that
this possibility could cause a larger effect on $g-2$ and other observables from vacuum polarization corrections. 
\begin{table}[htbp]
\begin{center}
  \begin{tabular}{|c|c|c|}
    \hline
    $d,~\mu,~\Lambda_3$ & $a_\mu^\U$ & $ \delta a_\mu^\U $  \\ \hline
    $ 1.1, 1, 500 $ & $1.7\times 10^{-5} $ & $ -3.2\times 10^{-6} $ \\ \hline
    $ 1.1, 1, 1000 $ & $1.5\times 10^{-5} $ & $ -2.7\times 10^{-6} $ \\ \hline
    $ 1.1, 1, 2000 $ & $1.3\times 10^{-5} $ & $ -2.3\times 10^{-6} $ \\ \hline
    $ 1.1, 10, 500 $ & $2.9\times 10^{-7} $ & $ -1.\times 10^{-4} $ \\ \hline
    $ 1.1, 10, 1000 $ & $2.5\times 10^{-7} $ & $ -8.8\times 10^{-5} $ \\ \hline
    $ 1.1, 10, 2000 $ & $2.2\times 10^{-7} $ & $ -7.6\times 10^{-5} $ \\ \hline
    $ 1.1, 100, 500 $ & $4.6\times 10^{-9} $ & $ -4.1\times 10^{-5} $ \\ \hline
    $ 1.1, 100, 1000 $ & $4.\times 10^{-9} $ & $ -2.3\times 10^{-5} $ \\ \hline
    $ 1.1, 100, 2000 $ & $3.5\times 10^{-9} $ & $ -1.3\times 10^{-5} $ \\ \hline
    $ 1.3, 1, 500 $ & $6.7\times 10^{-7} $ & $ -1.3\times 10^{-5} $ \\ \hline
    $ 1.3, 1, 1000 $ & $4.4\times 10^{-7} $ & $ -5.2\times 10^{-6} $ \\ \hline
    $ 1.3, 1, 2000 $ & $2.9\times 10^{-7} $ & $ -1.9\times 10^{-6} $ \\ \hline
    $ 1.3, 10, 500 $ & $2.9\times 10^{-8} $ & $ -2.\times 10^{-4} $ \\ \hline
    $ 1.3, 10, 1000 $ & $1.9\times 10^{-8} $ & $ -9.8\times 10^{-5} $ \\ \hline
    $ 1.3, 10, 2000 $ & $1.2\times 10^{-8} $ & $ -4.7\times 10^{-5} $ \\ \hline
    $ 1.3, 100, 500 $ & $1.1\times 10^{-9} $ & $ -2.\times 10^{-3} $ \\ \hline
    $ 1.3, 100, 1000 $ & $7.5\times 10^{-10} $ & $ -1.\times 10^{-3} $ \\ \hline
    $ 1.3, 100, 2000 $ & $4.9\times 10^{-10} $ & $ -5.\times 10^{-4} $ \\ \hline
    $ 1.5, 1, 500 $ & $2.9\times 10^{-8} $ & $ -4.2\times 10^{-4} $ \\ \hline
    $ 1.5, 1, 1000 $ & $1.4\times 10^{-8} $ & $ -2.3\times 10^{-4} $ \\ \hline
    $ 1.5, 1, 2000 $ & $7.1\times 10^{-9} $ & $ -1.2\times 10^{-4} $ \\ \hline
    $ 1.5, 10, 500 $ & $3.\times 10^{-9} $ & $ -2.8\times 10^{-3} $ \\ \hline
    $ 1.5, 10, 1000 $ & $1.5\times 10^{-9} $ & $ -1.6\times 10^{-3} $ \\ \hline
    $ 1.5, 10, 2000 $ & $7.5\times 10^{-10} $ & $ -8.8\times 10^{-4} $ \\ \hline
    $ 1.5, 100, 500 $ & $3.\times 10^{-10} $ & $ -1.7\times 10^{-2} $ \\ \hline
    $ 1.5, 100, 1000 $ & $1.5\times 10^{-10} $ & $ -1.\times 10^{-2} $ \\ \hline
    $ 1.5, 100, 2000 $ & $7.5\times 10^{-11} $ & $ -5.9\times 10^{-3} $ \\ \hline
  \end{tabular}
\end{center}
\caption{The relative difference $\delta(a_\mu^\U)$ between the unparticle contribution to muon $g-2$ with and without
  vacuum polarization corrections to the unparticle propagator. 
  An entry in the first column gives the choice of parameters used for that row, with $\mu$ and $\Lambda_3$ given in GeV.
  The second column gives
  $a_\mu^\U$ from unparticles with no vacuum polarization corrections.  The remaining 
  column gives the fractional difference $\delta(a_\mu^\U)=(\tilde{a}_\mu^\U-a_\mu^\U)/(a_\mu^\U)$, where $\tilde{a}_\mu^\U$
  takes into account vacuum polarization correction to the unparticle propagator from all charged SM fermions.}
\label{tbltwo}
\end{table}

Strong experimental constraints exist on $g-2$ for both muons and electrons.  In the case of muons, the measured value 
of $a$ is three standard deviations larger than the calculated SM value.  Specifically \cite{Bennett:2006fi}, 
\begin{equation}
  a^{\rm exp}_\mu-a^{\rm SM}_\mu=2.95\pm 0.81 \times 10^{-9}\;.
\end{equation}
As a very conservative upper bound, we require that the unparticle contribution, which is always positive in the case of
vector unparticles, be no more than three standard
deviations above the central value, i.e.
\begin{equation}
  a_\mu^\U<5.38\times 10^{-9}\;.
\end{equation}

In the case of electrons, the measured and calculated values are in good agreement.  We will take the conservative bound
\cite{Liao:2007bx} $a_e^\U<1.5\times 10^{-11}$, which, we note, is over five times the error on the most recent
experimental measurement \cite{Hanneke:2008tm}.

\begin{figure}%
  \centering%
  \subfloat[][]{%
    \includegraphics[width=5.5cm]{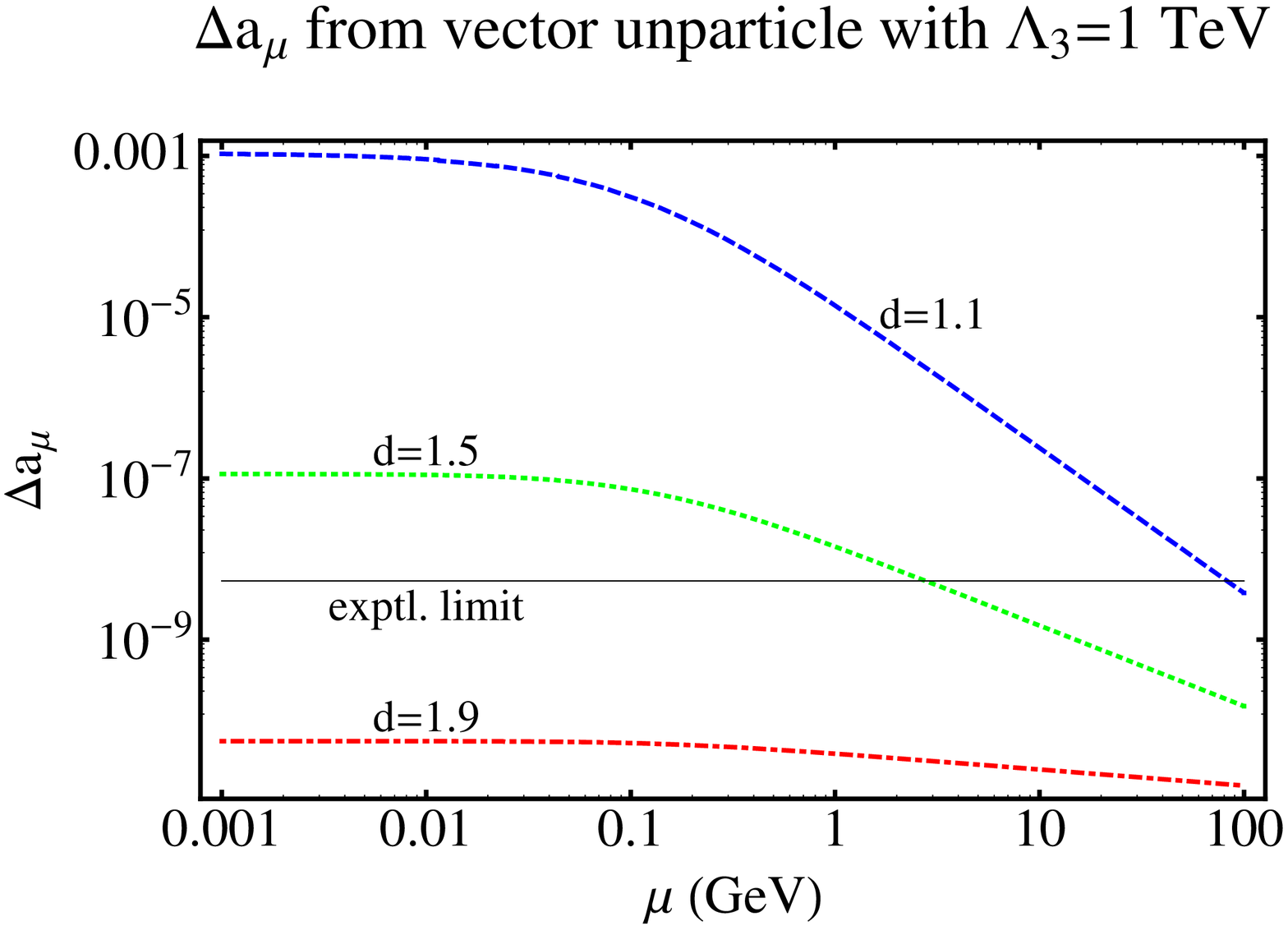}%
    \label{fig:pamuvsmu}%
  }%
  \qquad%
  \subfloat[][]{%
    \includegraphics[width=5.5cm]{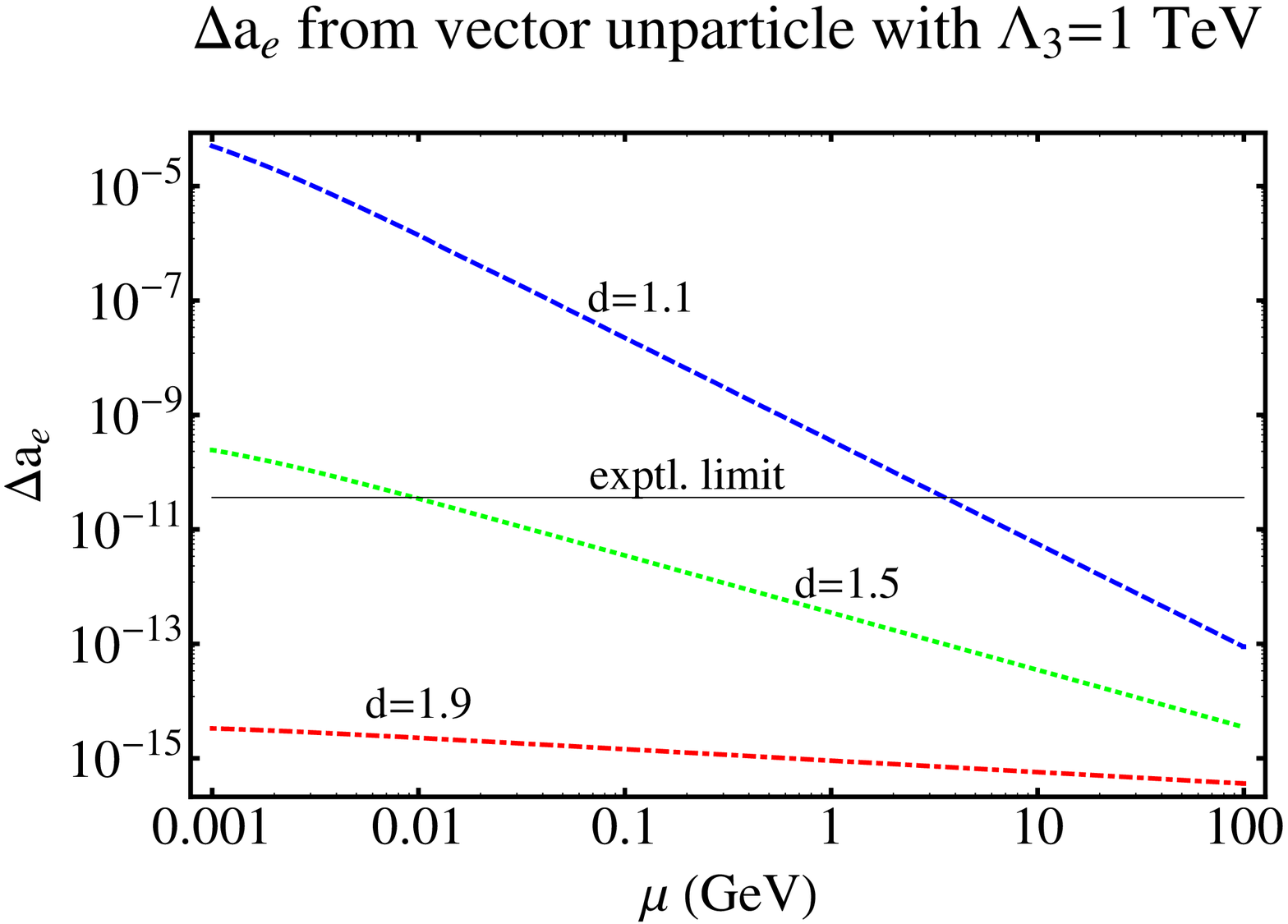}%
    \label{fig:paevsmu}%
  }%
  \caption{The unparticle contribution to the muon anomalous magnetic moment ($a_\mu^\U$) (left panel),
    and to the electron anomalous magnetic moment  ($a_e^\U$) (right panel),
    as a function of the conformal breaking scale $\mu$.  Here, $\Lambda_3=1~\tev$ and we show curves for 
    $d=1.1$, $d=1.5$, and $d=1.9$.
    Shown for comparison in each case is the experimental limit (see text).}%
  \label{fig:1figs}%
\end{figure}

In Fig.~\ref{fig:pamuvsmu}, we display the unparticle contribution to muon $g-2$ in the case of broken conformal
symmetry as a function of $\mu$ and compare it to the experimental bounds.  As described above, this contribution is 
obtained by numerically integrating Eq.~\ref{dispa}, again using {\it Mathematica}.  The contribution begins to
drop quickly as $\mu$ is increased above $m_\mu$, and the statement that for $\mu\gtrsim 1~\gev$, the experimental
bound is evaded, is shown to be true for, e.g., $\Lambda_3=1~\tev$ and $d=1.5$.  It is interesting to note the
large hierarchy between the $d=1.1,~1.5$, and $1.9$ contributions.



\begin{figure}%
  \centering
  \subfloat[][]{%
    \includegraphics[width=5.5cm]{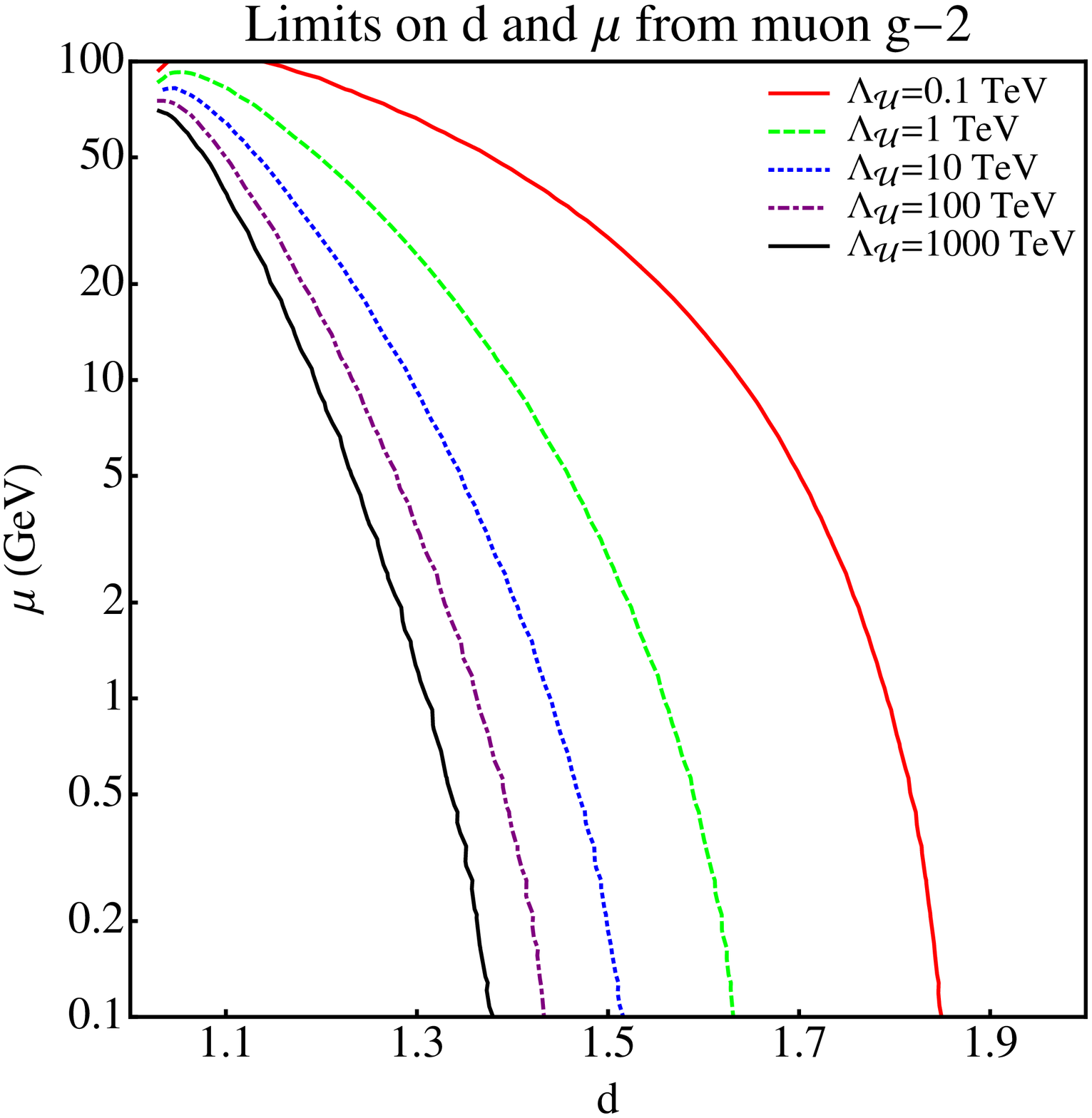}%
    \label{fig:dmuboundmu}%
  }%
  \qquad
  \subfloat[][]{%
    \includegraphics[width=5.5cm]{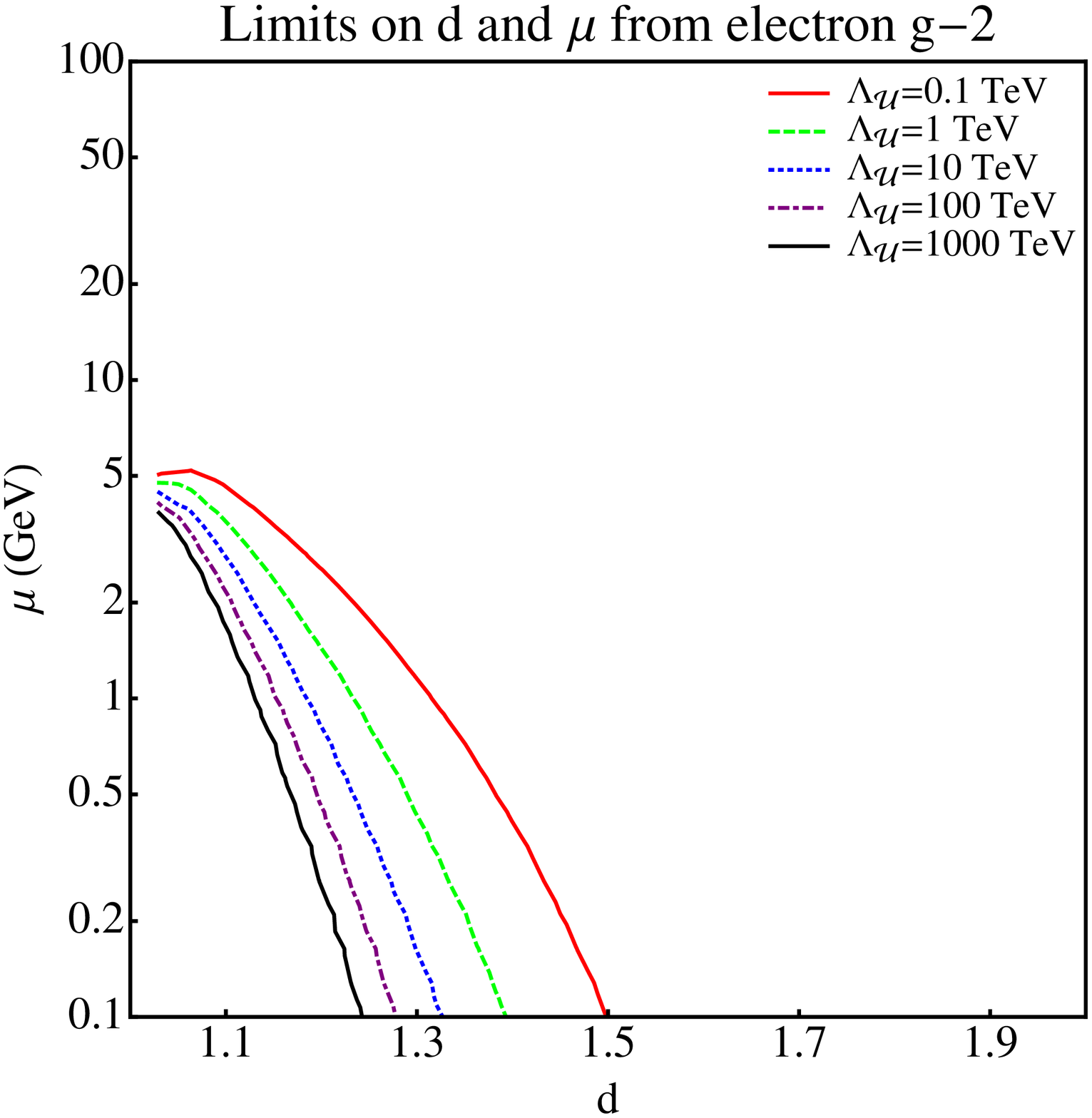}%
    \label{fig:dmubounde}%
  }%
  \caption{Lower bounds on $d$ and $\mu$ from muon (left panel) or electron (right panel)
    $g-2$ for various values of $\Lambda_3$.  The region above the 
    curves is allowed.}%
\end{figure}

Finally, in Fig.~\ref{fig:dmuboundmu}, we show the lower bounds on $d$ and $\mu$ from
muon $g-2$ for various choices of $\Lambda_3$.  We note the very strong dependence on $d$ and $\mu$.
In Fig.~\ref{fig:dmubounde}, we display the same bounds but from electron $g-2$.  It is qualitatively similar
though the bounds are much weaker.


\section{Conclusions}

In this paper we have examined the correction to lepton $g-2$ in
several unparticle scenarios.  
In particular, we studied the model in which in addition to breaking
the conformal symmetry of the unparticle sector at a scale $\mu$, we
also added a width-like term from vacuum polarization to the
propagator.  
To calculate the effects of this term we utilized an approximation based on 
dispersion integral techniques, which is exact when there are no
vacuum polarization corrections to the propagator.  
This approximation is accurate for $d \leq 1.5$; it breaks down
for greater values of $d$.

We found that adding the width-like term to the propagator
significantly changed the the value of the propagator, but made little
difference in the calculation of $g-2$.  Specifically, for the points
we analyzed numerically using our dispersion integral approximation,
we found that adding this term never changed the value of $g-2$ by
more than 1\%.  It would be interesting to learn whether this is the
case for other observables or whether $g-2$ is exceptionally
insensitive to this term in the propagator.

We also obtained limits on the unparticle dimension $d$ and the conformal scale $\mu$
for various values of the unparticle scale $\Lambda_3$
by assuming that the dimensionless coupling $c$ is equal to unity.  In
the muon case, we find that even for $\Lambda_3$ as high as 1000~\tev, 
there are strong bounds on $\mu$ for $d\lesssim1.3$.  In the electron 
case, the bounds are much weaker, though we still find that for
$\Lambda_3=1000~\tev$, $\mu\gtrsim 1~\gev$ for $d\lesssim 1.2$.


\section*{Acknowledgements}
We would like to thank J. Hewett and T. Rizzo for useful discussions regarding this work.
This work was supported by the US Department of Energy, contract DE--AC02--76SF00515.

\end{fmffile}

\bibliography{allUnTown2}{}
\bibliographystyle{h-elsevier3}
  
\end{document}